\documentclass[]{article}
\usepackage{lmodern}
\usepackage{amssymb,amsmath}
\usepackage{ifxetex,ifluatex}
\usepackage{fixltx2e} 
\ifnum 0\ifxetex 1\fi\ifluatex 1\fi=0 
  \usepackage[T1]{fontenc}
  \usepackage[utf8]{inputenc}
\else 
  \ifxetex
    \usepackage{mathspec}
    \usepackage{xltxtra,xunicode}
  \else
    \usepackage{fontspec}
  \fi
  \defaultfontfeatures{Mapping=tex-text,Scale=MatchLowercase}
  
\fi
\IfFileExists{upquote.sty}{\usepackage{upquote}}{}
\IfFileExists{microtype.sty}{%
\usepackage{microtype}
\UseMicrotypeSet[protrusion]{basicmath} 
}{}
\usepackage[margin=1in]{geometry}
\ifxetex
  \usepackage[setpagesize=false, 
              unicode=false, 
              xetex]{hyperref}
\else
  \usepackage[unicode=true]{hyperref}
\fi
\hypersetup{breaklinks=true,
            bookmarks=true,
            pdfauthor={John J. Nay, Yevgeniy Vorobeychik},
            pdftitle={Predicting Human Cooperation},
            colorlinks=true,
            citecolor=blue,
            urlcolor=blue,
            linkcolor=magenta,
            pdfborder={0 0 0}}
\usepackage{longtable,booktabs}
\usepackage{graphicx,grffile}
\makeatletter
\def\maxwidth{\ifdim\Gin@nat@width>\linewidth\linewidth\else\Gin@nat@width\fi}
\def\maxheight{\ifdim\Gin@nat@height>\textheight\textheight\else\Gin@nat@height\fi}
\makeatother
\setkeys{Gin}{width=\maxwidth,height=\maxheight,keepaspectratio}
\let\rmarkdownfootnote\footnote%
\def\footnote{\protect\rmarkdownfootnote}

\usepackage{titling}

  \title{Predicting Human Cooperation}
  \pretitle{\vspace{\droptitle}\centering\huge}
  \posttitle{\par}
  \author{John J. Nay, Yevgeniy Vorobeychik}
  \preauthor{\centering\large\emph}
  \postauthor{\par}
  \predate{\centering\large\emph}
  \postdate{\par}
  \date{April 5, 2016}

\ifx\paragraph\undefined\else
\let\oldparagraph\paragraph
\renewcommand{\paragraph}[1]{\oldparagraph{#1}\mbox{}}
\fi
\ifx\subparagraph\undefined\else
\let\oldsubparagraph\subparagraph
\renewcommand{\subparagraph}[1]{\oldsubparagraph{#1}\mbox{}}
\fi

\begin{document}
\maketitle

\section{Abstract}\label{abstract}

The Prisoner's Dilemma has been a subject of extensive research due to
its importance in understanding the ever-present tension between
individual self-interest and social benefit. A strictly dominant
strategy in a Prisoner's Dilemma (defection), when played by both
players, is mutually harmful. Repetition of the Prisoner's Dilemma can
give rise to cooperation as an equilibrium, but defection is as well,
and this ambiguity is difficult to resolve. The numerous behavioral
experiments investigating the Prisoner's Dilemma highlight that players
often cooperate, but the level of cooperation varies significantly with
the specifics of the experimental predicament. We present the first
computational model of human behavior in repeated Prisoner's Dilemma
games that unifies the diversity of experimental observations in a
systematic and quantitatively reliable manner. Our model relies on data
we integrated from many experiments, comprising 168,386 individual
decisions. The computational model is composed of two pieces: the first
predicts the first-period action using solely the structural game
parameters, while the second predicts dynamic actions using both game
parameters and history of play. Our model is extremely successful not
merely at fitting the data, but in predicting behavior at multiple
scales in experimental designs not used for calibration, using only
information about the game structure. We demonstrate the power of our
approach through a simulation analysis revealing how to best promote
human cooperation.

\textbf{Keywords}: Social dilemma; Prisoner's Dilemma; Repeated games;
Predictive modeling; Computer simulation; Institutional design

\footnotetext[1]{See http://johnjnay.com/ for author contact details.}

\section{Introduction}\label{introduction}

The Prisoner's Dilemma game has been a subject of extensive research due
to its importance in understanding the ever-present tension between
individual self-interest and social benefit {[}1--3{]}. From a
theoretical perspective, a strictly dominant strategy (defection), when
played by both players, is mutually harmful: cooperation by both yields
significant mutual benefits relative to defection. For example, local
maintenance of shared drinking water systems in rural communities
represents a Prisoner's Dilemma that can result in a ``tragedy of the
commons'' {[}4{]}. From each community member's perspective, they are
better off if someone else invests in maintaining the infrastructure. If
the majority of the community adopts this strategy, everyone is worse
off because the system breaks down and no longer provides clean water.

In most social dilemma settings, however, interactions are repeated.
Thus, for example, community members must repeatedly make water
infrastructure investment decisions. Repetition of the Prisoner's
Dilemma, a more realistic model of human interaction than a one-shot
game, can theoretically give rise to cooperation as an equilibrium if
players are sufficiently patient; still, defection remains an
equilibrium as well, and this ambiguity is difficult to resolve. In
particular, theoretical treatment of repeated Prisoner's Dilemma games
is not instructive in identifying when cooperation or defection emerges
as the predominant outcome. Given the limitations of theory in
explaining repeated cooperation, researchers have turned to experiments
to better understand behavior and the effects of institutional structure
on social outcome by considering different game structures and
investigating associated cooperation proclivities of human subjects
{[}5{]}. The experiments highlight that humans often cooperate, but the
overall level and temporal evolution of cooperation vary significantly
with the specific design.

We develop a predictive model of dynamic cooperation that reliably
forecasts behavior across heterogeneous game designs, and then analyze
this model to tease apart the magnitude and direction of the effects of
game design variables on cooperation. For this purpose we compiled data
from previously analyzed repeated Prisoner's Dilemma experiments
{[}6--13{]}. We created standardized measures of the game and individual
behavior across these games, and used machine learning techniques to
calibrate and evaluate computational models. Our model is extremely
successful in predicting individual decisions, average cooperation
levels, and cooperation dynamics \emph{in games not used for model
calibration}. Moreover, we demonstrate that this synthetic model can
predict the high-level quantitative and qualitative findings of the
human subject experiments.

The long-term goal of this research program is to map the experimental
variables onto real-world policy design factors and use model analyses
to inform policies that facilitate cooperation where the underlying
social structure would otherwise lead to a breakdown. For instance, how
can we best design development programs that lead to sufficient
voluntary maintenance of shared water systems? Is it more important to
increase the potential benefits of mutual cooperation over mutual
defection, or to increase the benefits of mutual cooperation over losing
out by being the sole cooperator?

\section{Data}\label{data}

The data are from human subjects experiments that used real financial
incentives and transparently conveyed the rules of the game to the
subjects, which is standard procedure in experimental economics.
Subjects anonymously interact and their decisions to cooperate or defect
at each time period of each interaction are recorded. They receive
payoffs proportional to the outcomes in a specified payoff table similar
to Table 1. From the description of the experiments in the published
papers and the publicly available data sets, we were able to build a
comprehensive collection of game structures and individual decisions.

The thirty game structures that we compiled varied substantially across
a number of dimensions, aside from player payoffs. In some structures,
payoffs were deterministic, whereas others featured stochastic payoffs
(in this case, the expected payoffs constituted the payoff structure).
In some structures, players imperfectly observed their counterparts'
past actions. Another key distinction was whether or not a game had a
fixed time horizon, or would terminate independently after each
iteration with a fixed probability. Finally, while most games were
played over a discrete sequence of iterations, some were in continuous
time. We use nine variables to quantify game structure along these
salient dimensions. \emph{Risk} is an indicator of whether there is
stochasticity in the payoffs {[}8,10{]}. \emph{Error} is the probability
that the choice a player makes will be exogenously flipped {[}13{]}.
\emph{Infinite} is an indicator of whether interactions are indefinitely
repeated or have a fixed length {[}7{]}. \(\delta\) is the probability
that the next period of the current paired interaction will occur in a
infinitely game {[}11{]}. We used a formula,
\(E[Interaction Length] = \frac{1}{1 - \delta}\), to compute \(\delta\)
for finitely repeated interactions; for instance, the finitely repeated
interactions in {[}10{]} were all ten periods long so \(\delta = 0.9\).
\emph{Continuous} is an indicator of whether interactions are played in
``continuous time,'' rather than the standard discrete rounds {[}12{]}.
\emph{R} is the reward received if both players cooperate; \emph{P} is
the punishment received if both defect; \emph{T} is the temptation to
defect on the other; and \emph{S} is the payoff for being a sucker by
cooperating as the other defects (Table 1 illustrates the way in which
the four payoff values map onto the Prisoner's Dilemma bi-matrix
representation).

\begin{longtable}[c]{@{}lcc@{}}
\toprule
\begin{minipage}[b]{0.10\columnwidth}\raggedright\strut
~
\strut\end{minipage} &
\begin{minipage}[b]{0.10\columnwidth}\centering\strut
C
\strut\end{minipage} &
\begin{minipage}[b]{0.10\columnwidth}\centering\strut
D
\strut\end{minipage}\tabularnewline
\midrule
\endhead
\begin{minipage}[t]{0.10\columnwidth}\raggedright\strut
\textbf{C}
\strut\end{minipage} &
\begin{minipage}[t]{0.10\columnwidth}\centering\strut
\emph{(R,R)}
\strut\end{minipage} &
\begin{minipage}[t]{0.10\columnwidth}\centering\strut
\emph{(S,T)}
\strut\end{minipage}\tabularnewline
\begin{minipage}[t]{0.10\columnwidth}\raggedright\strut
\textbf{D}
\strut\end{minipage} &
\begin{minipage}[t]{0.10\columnwidth}\centering\strut
\emph{(T,S)}
\strut\end{minipage} &
\begin{minipage}[t]{0.10\columnwidth}\centering\strut
\emph{(P,P)}
\strut\end{minipage}\tabularnewline
\bottomrule
\end{longtable}

\textbf{Table 1.} Payoff table where one player plays from the
perspective of the columns and the other from the rows. For this to be a
repeated Prisoner's Dilemma, it must hold that \(T>R>P>S\), and
\(R > (S + T)/2\) {[}14{]}.

To create standardized payoff measures from the \emph{R, S, T, P}
values, we used two differences between payoffs associated with
important game outcomes, both normalized by the difference between the
temptation to defect and being a sucker when cooperating as the other
defects {[}15{]}. \(r_1\) is the normalized difference between the
reward received if both players cooperate and the punishment received if
both defect, \(\frac{R-P}{T-S}\). \(r_2\) is the normalized difference
between the reward received if both players cooperate and the payoff for
being a sucker when cooperating as the other defects,
\(\frac{R-S}{T-S}\). Because \(\frac{\partial r1}{\partial R} > 0\),
\(\frac{\partial r1}{\partial P} < 0\),
\(\frac{\partial r1}{\partial T} < 0\), and
\(\frac{\partial r1}{\partial S} > 0\), \(r_1\) has been used as an
index of the cooperativeness of a Prisoner's Dilemma {[}15,16{]}, while
\(r_2\) is descriptive of how much better off a player will be if their
opponent cooperates, rather than defects, while they themselves
cooperate. Table 2 summarizes the game structures from the data sets we
standardized and combined.

\begin{longtable}[c]{@{}rrrrrrrrl@{}}
\toprule
Error & Delta & Infinity & Continuous & Risk & r1 & r2 & Cooperation &
Dataset\tabularnewline
\midrule
\endhead
0.0000 & 0.900 & 0 & 0 & 0 & 0.18 & 0.590 & 0.60 & BR\tabularnewline
0.0000 & 0.900 & 0 & 0 & 1 & 0.18 & 0.590 & 0.35 & BR\tabularnewline
0.0000 & 0.900 & 1 & 0 & 0 & 0.33 & 0.670 & 0.56 & DO\tabularnewline
0.0000 & 0.900 & 0 & 0 & 1 & 0.33 & 0.830 & 0.31 & KS\tabularnewline
0.0000 & 0.900 & 0 & 0 & 0 & 0.33 & 0.830 & 0.57 & KS\tabularnewline
0.0000 & 0.500 & 1 & 0 & 0 & 0.18 & 0.530 & 0.10 & DF\tabularnewline
0.0000 & 0.750 & 1 & 0 & 0 & 0.18 & 0.530 & 0.20 & DF\tabularnewline
0.0000 & 0.500 & 1 & 0 & 0 & 0.39 & 0.740 & 0.18 & DF\tabularnewline
0.0000 & 0.750 & 1 & 0 & 0 & 0.39 & 0.740 & 0.59 & DF\tabularnewline
0.0000 & 0.750 & 1 & 0 & 0 & 0.61 & 0.950 & 0.76 & DF\tabularnewline
0.0000 & 0.500 & 1 & 0 & 0 & 0.61 & 0.950 & 0.35 & DF\tabularnewline
0.1250 & 0.875 & 1 & 0 & 0 & 0.20 & 0.600 & 0.34 & FR\tabularnewline
0.1250 & 0.875 & 1 & 0 & 0 & 0.33 & 0.660 & 0.49 & FR\tabularnewline
0.1250 & 0.875 & 1 & 0 & 0 & 0.43 & 0.710 & 0.59 & FR\tabularnewline
0.0000 & 0.875 & 1 & 0 & 0 & 0.60 & 0.800 & 0.74 & FR\tabularnewline
0.0625 & 0.875 & 1 & 0 & 0 & 0.60 & 0.800 & 0.78 & FR\tabularnewline
0.1250 & 0.875 & 1 & 0 & 0 & 0.60 & 0.800 & 0.57 & FR\tabularnewline
0.0000 & 0.900 & 0 & 0 & 0 & 0.25 & 0.583 & 0.43 & AM\tabularnewline
0.0000 & 0.875 & 0 & 1 & 0 & 0.11 & 0.560 & 0.27 & FO\tabularnewline
0.0000 & 0.875 & 0 & 1 & 0 & 0.14 & 0.710 & 0.33 & FO\tabularnewline
0.0000 & 0.875 & 0 & 1 & 0 & 0.33 & 0.560 & 0.54 & FO\tabularnewline
0.0000 & 0.875 & 0 & 1 & 0 & 0.43 & 0.710 & 0.62 & FO\tabularnewline
0.0000 & 0.500 & 0 & 0 & 0 & 0.33 & 0.610 & 0.12 & DB\tabularnewline
0.0000 & 0.750 & 0 & 0 & 0 & 0.33 & 0.610 & 0.24 & DB\tabularnewline
0.0000 & 0.750 & 0 & 0 & 0 & 0.33 & 0.720 & 0.25 & DB\tabularnewline
0.0000 & 0.500 & 0 & 0 & 0 & 0.33 & 0.720 & 0.13 & DB\tabularnewline
0.0000 & 0.500 & 1 & 0 & 0 & 0.33 & 0.610 & 0.23 & DB\tabularnewline
0.0000 & 0.750 & 1 & 0 & 0 & 0.33 & 0.610 & 0.35 & DB\tabularnewline
0.0000 & 0.750 & 1 & 0 & 0 & 0.33 & 0.720 & 0.36 & DB\tabularnewline
0.0000 & 0.500 & 1 & 0 & 0 & 0.33 & 0.720 & 0.31 & DB\tabularnewline
\bottomrule
\end{longtable}

\textbf{Table 2.} Summary of thirty game structures that compose the
full combined data set {[}6--13{]}. BR 2006 {[}8{]} and DB 2005 {[}7{]}
both also conducted one-shot games; we only describe and use their
repeated game data. KSBS 2009 {[}10{]} also conducted games with partial
information; we only describe and use their full information data. AM
1993 {[}6{]} also conducted games that matched humans with computers; we
only describe and use the games they conducted where humans played other
humans. FO 2012 {[}12{]} included one-shot games and games with very
different protocols for how and when to make a choice in order to study
continuous choices; we only use the ``Grid treatment with n = 8
subperiods,'' which they say is, ``comparable to the 10-stage repeated
games featured in previous laboratory studies.'' DO 2009 {[}9{]} also
conducted random matching of opponents; we only use their fixed matching
treatments.

Fig. 1 plots game structures based on their values of the four
quantitative game structure variables, \(r_1\), \(r_2\), \(\delta\), and
\emph{error}, illustrating the broad empirical support in our combined
data across the values of these variables, and that there are no
\emph{simple} relationships between these variables and the proportion
of cooperation that can be detected without, at least, controlling for
variables not included in each plot.

\includegraphics{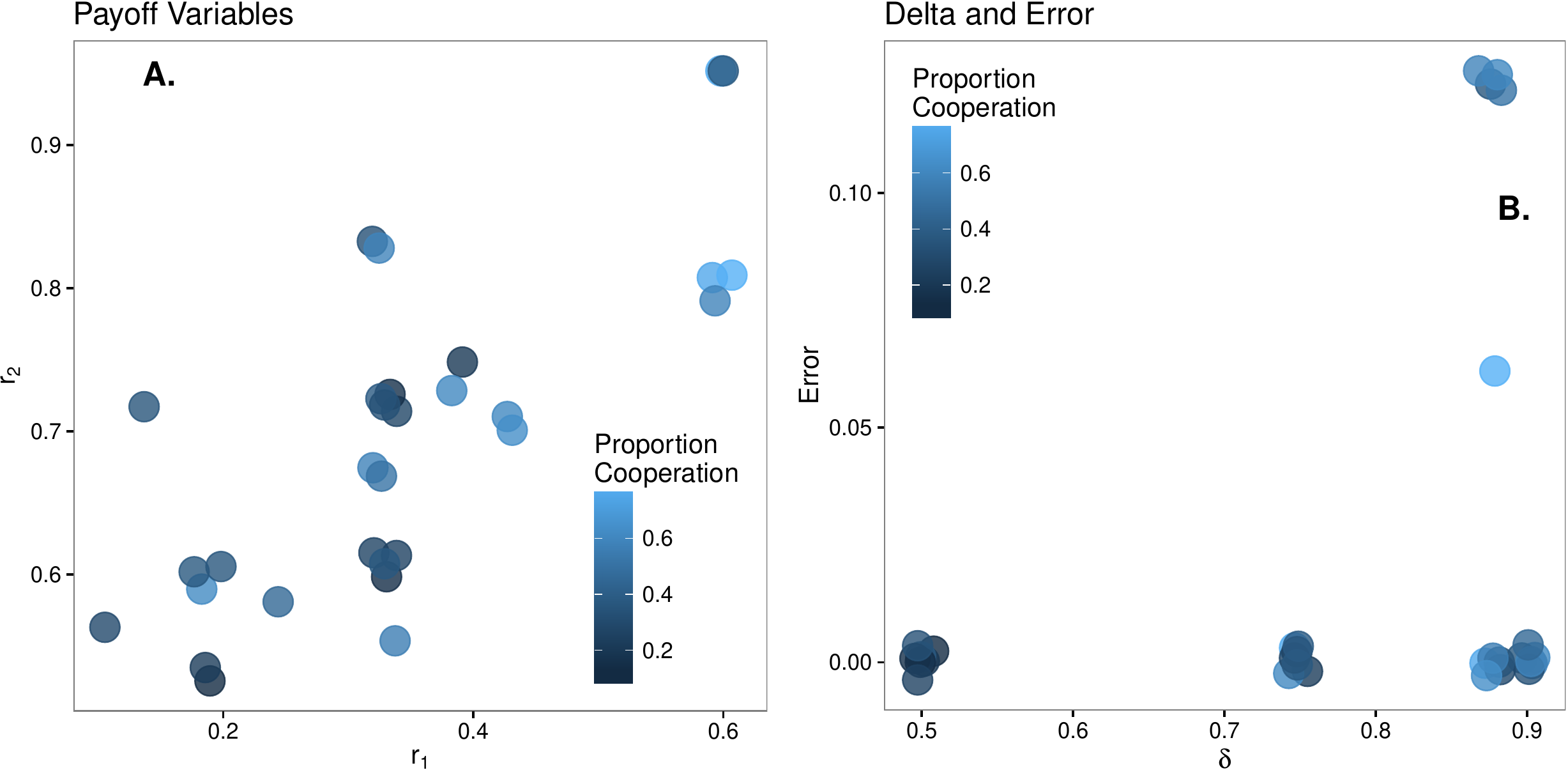}

\textbf{Fig. 1.} Game structures (\emph{n = 30}) with location based on
the payoff variable values (\textbf{A}.), and delta and error values
(\textbf{B}.). Colors represent proportion of cooperation observed in
the game structure. Locations have been slightly randomly shifted to
improve visualization.

Our combined data set can be organized hierarchically (Fig. 2). Within
each game structure, there are interactions between pairs of players;
these are repetitions of the same ``stage-game'' between the same two
players. Repeating the game with past behavior as common knowledge can
theoretically increase cooperation by bringing players' reputational
concerns into play. Within each interaction, there are time periods.
Finally, in each time period, both players simultaneously take a single
action.

\includegraphics{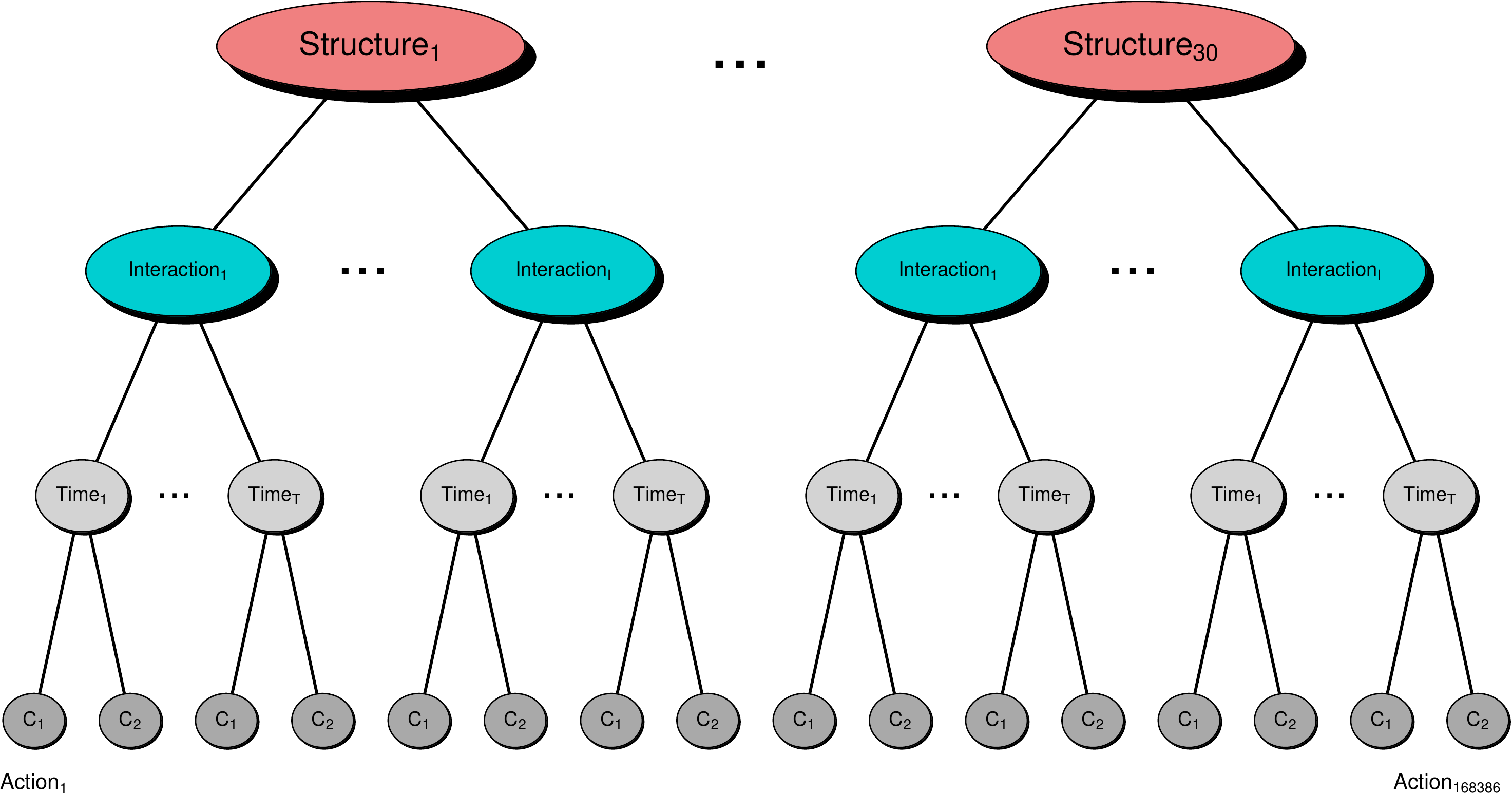}

\textbf{Fig. 2.} A hierarchical view of our data. The top level divides
the data into 30 game structures. The next level down are the
interactions between two players. Within each interaction, there are
\emph{T} time periods. In repeated games in which termination is
stochastic, \emph{T} ranges up to thirty-eight. Across all interactions
and structures, \emph{T} is five, on average. Within each time period,
player 1 takes action \(C_1\) and player 2 takes action \(C_2\). 168,386
actions were taken across all the experimental data.

Our goal was to predict behavioral patterns simultaneously at several
levels within this hierarchy. Specifically, we wish to predict the
effects of the game structure on average cooperation (the highest level
in the hierarchy), the temporal dynamics of cooperation as a function of
structure (second lowest level), and individual-level actions (lowest
level). The impact of structure on cooperation has been the primary
subject of experimental investigations, with the natural goal of
understanding how to design institutions that promote cooperation.
Understanding both short-term and long-term impacts of institutions,
however, necessitates looking at behavior dynamics, rather than simply
aggregate levels of cooperation. Indeed, cooperation may well be high
early, but degrade with time, particularly close to the final period of
the game, if it is known {[}8{]}. Finally, understanding individual
behavior enables us to understand aggregate cooperation dynamics in
terms of micro decision processes. If our computational model can
successfully predict behavior at all levels in the hierarchy, we can
have confidence in the ability of the resulting model to generalize
experimental findings to new institutional structures, allowing us to
achieve the ultimate goal: a validated computational framework for
designing institutions that promote cooperation.

\section{Model}\label{model}

Our behavioral model has two parts: a ``static'' component that predicts
a player's first period action, and a ``dynamic'' component that
predicts a player's actions in subsequent times of the same interaction.
Both components are logistic regressions mapping a vector of predictor
variables into the probability of cooperation, with parameters learned
through maximum likelihood estimation on training data. The predictor
variables for first period play include the game structure,
\(\vec{Game}\), (\(r_1\), \(r_2\), \emph{risk}, \emph{error},
\(\delta\), \emph{infinite}, \emph{continuous}), and the predictors for
all other time periods (i.e., the dynamic model) include the game
structure, the actions of both players from the previous period,
\(\vec{History_{t-1}}\), and the current time period, \(t\). The
inclusion of the history of the interaction is motivated by evidence
that most participants in repeated cooperation games condition their
actions on previous play {[}17{]}. In mathematical terms, the
probability of cooperation, \(p(C_t)\), can be expressed as follows:

\[
p(C_t) = \left\{
     \begin{array}{lr}
       t=1 &  f_{static}(\vec{Game})\\
       t>1 &  f_{dynamic}(\vec{Game}, \vec{History_{t-1}}, t)
     \end{array}
   \right.
\]

where \emph{f} is determined by the logistic regression model (distinct
in both cases), calibrated on behavioral data.

Specifically, we used the following equation for first period
cooperation, \(C_{t=1}\):

\[r_1 + r_2 + risk + error + \delta + r_1 \times \delta + r_2 \times \delta + infinity + continuous\]

We used the following equation for cooperation in periods greater than
one, \(C_{t>1}\):

\[r_1 + r_2 + risk + error + \delta + r_1 \times \delta + r_2 \times \delta + infinity + continuous + \delta \times infinity + \]
\[my.decision_{t-1} + other.decision_{t-1} + error \times other.decision_{t-1} + t\]

Both equations use all the structural game features,
\(r_1 + r_2 + risk + error + \delta + r_1 \times \delta + r_2 \times \delta + infinity + continuous\),
and because we hypothesized that \(\delta\) and the payoff variables may
have difference effects depending on the values of the other, we
interacted them. For the dynamic model, we added an interaction term
between \(\delta\) and infinity to capture the different effect that
\(\delta\) may have when it actually determines the length of the game
probabilistically. In finite games, \(\delta\) represents a rational
expectation of the length of the game from a first period perspective.
In infinite games, \(\delta\) represents a rational expectation of the
length of the game for all periods. Therefore, \(\delta\) is used as a
feature in the model of first period play, and for all periods of play
beyond period one there is an interaction term that multiplies the
indicator variable for whether a game is infinite by the value of
\(\delta\). We interacted error with \(other.decision_{t-1}\) because
the greater the value of error, the less sure the player is of the
actual decision of the other player in the previous time period. The
model is then a logistic sigmoid function,
\(\sigma(w^TX) = \frac{1}{1 + exp(-w^TX)}\), acting on a linear function
of these features, \textbf{X}, with a vector of weights, \(\vec{w}\),
the length of the feature set. The computational implementation we used
was the base R ``stats::glm'' function and the caret ``train'' function
{[}18,19{]}.

We compare our model's performance to three alternative logistic
regression models: ``static-only,'' which just uses the first component
of the ``full'' model; ``dynamic-only,'' which just uses the second
component; and ``baseline,'' which uses the observed average level of
cooperation. Comparing the full model to its components allows us to
understand the relative contributions of the components to its
predictive power. We also compare our model to a state-of-the-art
behavioral game theory model designed for forecasting play in
out-of-sample games: functional experience-weighted attraction learning
(fEWA) {[}20{]}. The actions available to agent \(i\), which are indexed
by \(j\), are assumed to have numerical attractions for each time \(t\),
\(A^j_i(t)\), and fEWA updates the attractions based on functions of
\(i\)'s experience up to time \(t\) and the payoffs of the game (\(i\)'s
chosen strategy is \(s_i(t)\), \(i\)'s opponent's chosen strategy is
\(s_{-i}(t)\), \(i\)'s payoffs are \(\pi_i(s_i^j(t), s_{-i}(t))\), and
\(I\) yields \(I(x,y) = 0\) if \(x \neq y\), and \(I(x,y) = 1\) if
\(x = y\)).

\[  
A^j_i(t) = \frac{\phi_i(t) N(t-1) A^j_i(t-1) + [\delta_{ij}(t) + (1-\delta_{ij}(t))I(s^j_i, s_i(t))] \pi_i(s_i^j, s_{-i}(t))}{N(t-1) \phi_i(t) + 1}
\]

Then, attractions are mapped into probabilities of choosing Cooperate or
Defect the next time period with a logistic stochastic response function
(see \emph{S1 Appendix} for model details).

\[
P^j_i(t+1) = \frac{e^{\lambda A^j_i(t)}}{\sum_{k=1}^{m_i}  e^{\lambda A^k_i(t)}}
\]

In order to use the empirical models of individual behavior to predict
interactive outcomes of new experimental designs, we simulate
discrete-time dynamic systems comprised of autonomous decision
algorithms (agents) that interact with each other. This allows us to
simulate the play of an experiment without any behavioral data from that
experiment. Player behavior is endogenous to the simulation model, which
only needs to be initialized with a game structure specification. There
have been a number of studies using simulations to investigate
cooperation games {[}21--27{]}, and simulations have been used to inform
institutional design of strategic interactions more broadly
{[}28--30{]}. There has been research on cooperative equilibria models
for predicting aggregate cooperation patterns {[}31,32{]}, and a
significant amount of work on individual-level behavioral models
{[}33--40{]}. Our work diverges from most such research in three
respects: (i) agent behavior is derived solely from individual-level
empirical data, and (ii) we rigorously validate our model's ability to
predict behavior by measuring performance on many unseen game
structures. {[}27{]} also derive agent behavior solely from
individual-level game data. However, we utilize data from many more
experimental designs and from a different game.

\section{Results}\label{results}

\subsection{Individual-level
performance}\label{individual-level-performance}

Our first investigation evaluates models' ability to predict
individual-level actions. We divide game structures into training and
test groups, estimate the parameters in training game structures, and
then predict actions in held-out game structures, conditioning on the
game structure and the empirically observed actions of the previous
period (for periods greater than one). We repeatedly execute the
process, each time slightly changing the split of the data so each game
structure will be in the test data once (Fig. 3). The end results are
out-of-sample predictions of all actions in each game structure. To make
predictions with the dynamic-only model, which will have missing values
for the lagged action outcomes at period one, we draw cooperate/defect
actions with equal probability (corresponding, approximately, to average
cooperation/defection split over all game structures). When we instead
impute ``period zero'' outcomes as mutual cooperation the results are
qualitatively the same. For this test, we measure the log-likelihood of
the observed actions in the test data, given model predictions, which is
a statistically proper method for evaluating the quality of
probabilistic predictions. We also discretize model outputs into
Cooperate or Defect to measure accuracy, which is the proportion of
actions where the predicted probability of cooperation was above (below)
0.5 when the observed action was cooperation (defection).

\includegraphics{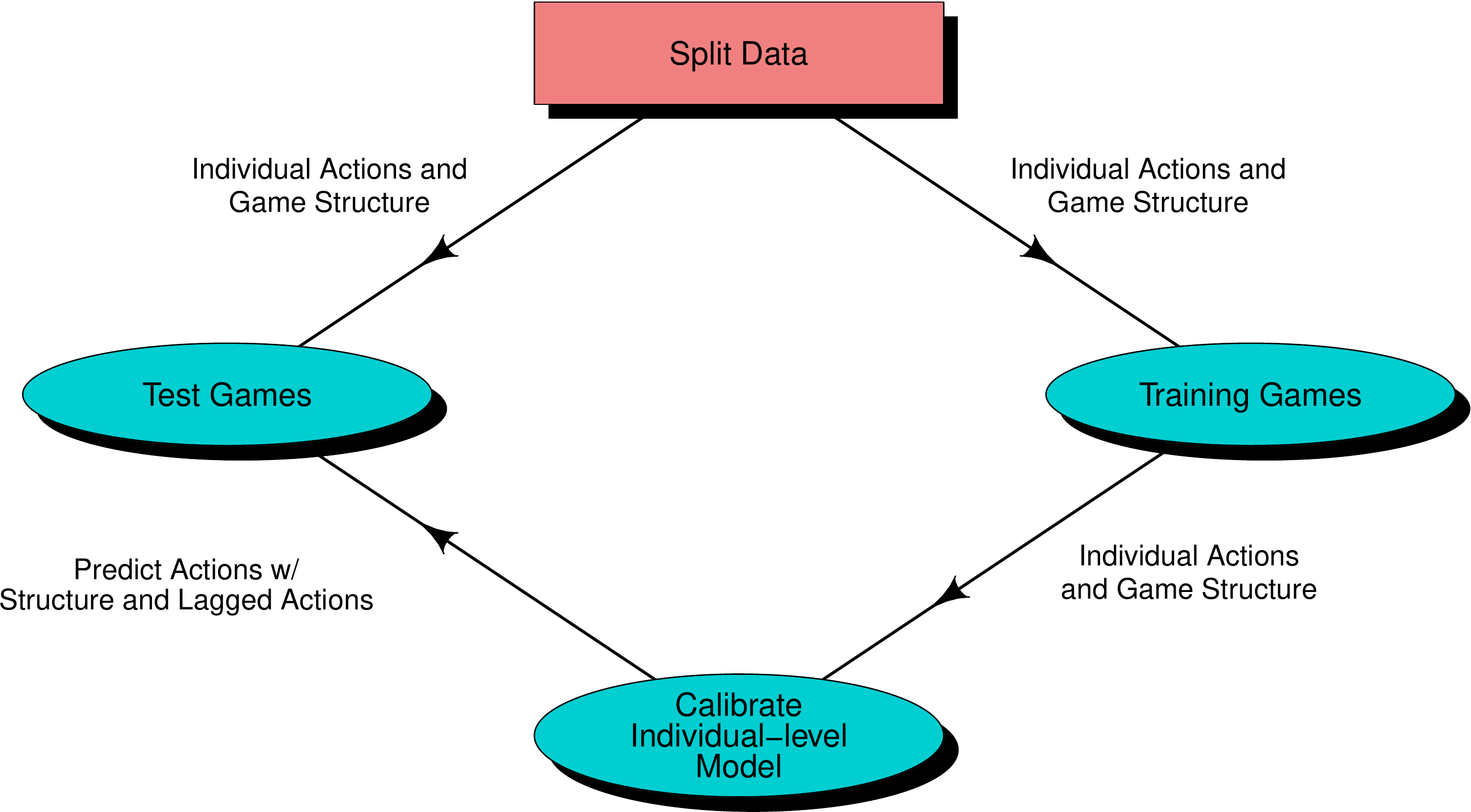}

\textbf{Fig. 3.} Model validation process for individual-level actions.
We assign each of the thirty game structures into either training or
test data. With the training data, we learn the parameters of the
individual-level model, and then predict the decisions in game
structures assigned to the test data. We repeat this process thirty
times, including a different game structure as the held-out test each
time (leave-out-one-cross-validation), until we have predictions for all
the decisions for each of the game structures.

The dynamic model performs almost as well as the full model in periods
greater than one, but poorly in the first period, indeed, worse than the
static model (Table 3). Overall, our relatively simple two-piece model
predicts the next action a player will take with 86\% accuracy on
average (a remarkably good prediction, given that human behavior is
generally quite noisy). Our model also significantly outperforms all
alternatives in terms of the log-likelihood measure, which is more
statistically appropriate in quantifying performance of stochastic
forecasts, but is less intuitive.

\subsection{Aggregate-level
performance}\label{aggregate-level-performance}

To evaluate the model's ability to predict behavior in new game
structures, we developed the following procedure (Fig. 4). Assign each
of the thirty game structures into either training or test data. With
the training data, learn the parameters of the individual-level model.
Next, create a simulation in which the estimated individual-level model
makes joint decisions in a repeated Prisoner's Dilemma game, and predict
the probabilistic behavior in game structures assigned to the test data
\emph{using only the game structure}, i.e., using no behavioral data
from the experiment. Finally, compare the predictions,
\(p(y_{sim}^{new} | \vec{Game}^{new})\), to actual observed cooperation
dynamics, \(y_{obs}^{new}\), using both squared error and correlation to
measure success of the model in predicting behavior. Repeat this process
thirty times, including a different game structure as the held-out test
each time (leave-out-one-cross-validation), until we have a prediction
for each of the game structures as if each prediction were made before
any data had been collected for that experimental design. We also test
that the results are robust to the number of folds in the
cross-validation procedure (from thirty down to three), i.e.~robust to
the number of game structures used for training.

\includegraphics{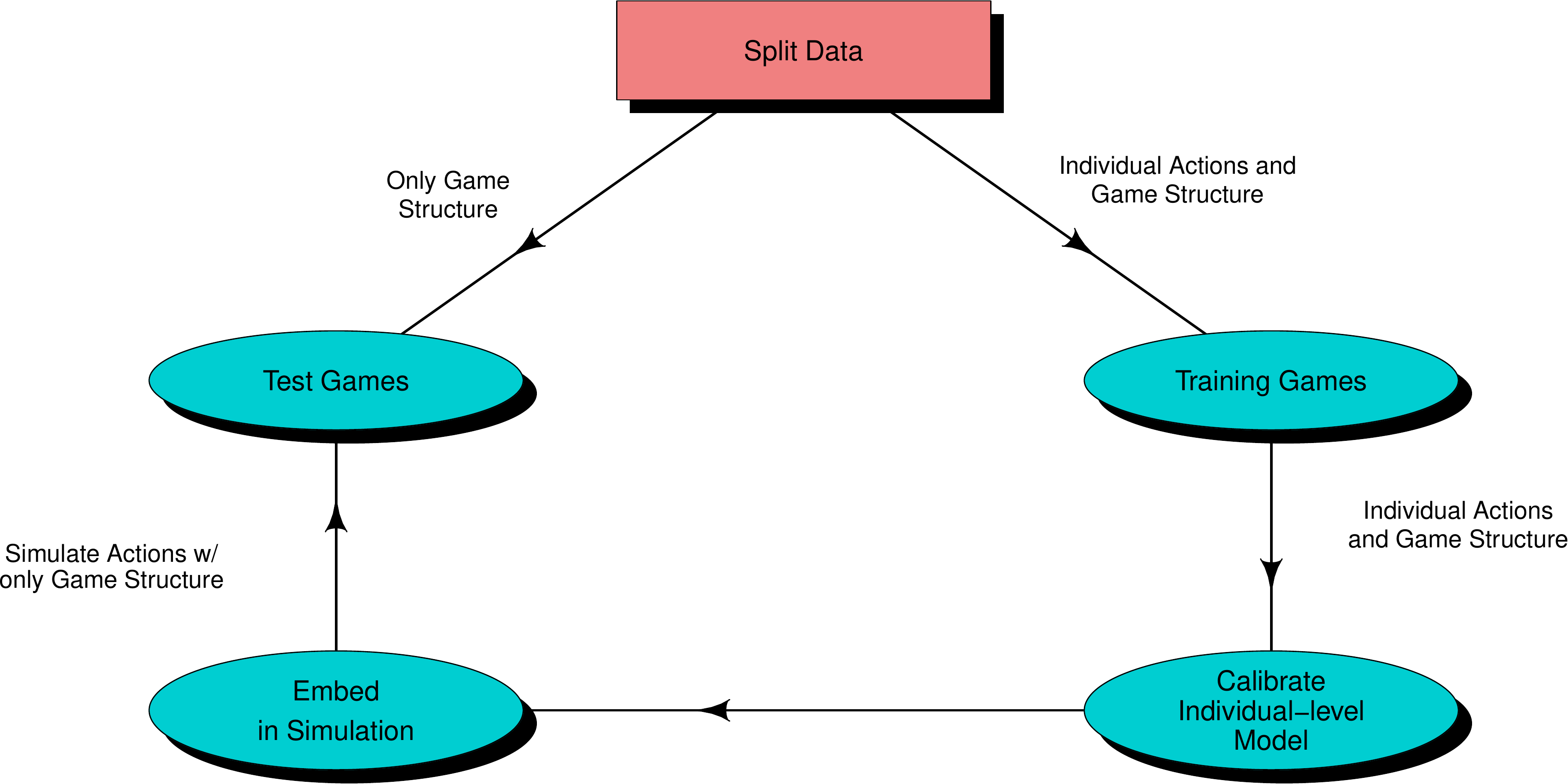}

\textbf{Fig. 4.} Model validation process for aggregate-level patterns.
We tested the dynamic-only model by sampling lagged outcomes for `period
zero' actions from a Bernoulli distribution with equal probability of
cooperation and defection, which is approximately the mean cooperation
rate in the data. A subtle, but crucial, distinction between this
process and the model validation process for individual-level action
predictions (Fig. 3) is that, here, \emph{we only pass game structures}
for the test games, rather than the full behavioral data and the game
stucture.

We compare the performance of the five models' predictions of average
probability of cooperation and dynamics of cooperation (Table 3). Our
model is slightly worse at predicting overall cooperation levels than
the static model, but better at predicting dynamics (neither comparison
is significant), and is significantly better than the other models in
almost all cases.

\begin{longtable}[c]{@{}cccccc@{}}
\toprule
\begin{minipage}[b]{0.19\columnwidth}\centering\strut
~
\strut\end{minipage} &
\begin{minipage}[b]{0.09\columnwidth}\centering\strut
Full
\strut\end{minipage} &
\begin{minipage}[b]{0.10\columnwidth}\centering\strut
Static
\strut\end{minipage} &
\begin{minipage}[b]{0.12\columnwidth}\centering\strut
Dynamic
\strut\end{minipage} &
\begin{minipage}[b]{0.08\columnwidth}\centering\strut
fEWA
\strut\end{minipage} &
\begin{minipage}[b]{0.12\columnwidth}\centering\strut
Baseline
\strut\end{minipage}\tabularnewline
\midrule
\endhead
\begin{minipage}[t]{0.19\columnwidth}\centering\strut
\textbf{Acc. t=1}
\strut\end{minipage} &
\begin{minipage}[t]{0.09\columnwidth}\centering\strut
\emph{68}
\strut\end{minipage} &
\begin{minipage}[t]{0.10\columnwidth}\centering\strut
62
\strut\end{minipage} &
\begin{minipage}[t]{0.12\columnwidth}\centering\strut
57
\strut\end{minipage} &
\begin{minipage}[t]{0.08\columnwidth}\centering\strut
48
\strut\end{minipage} &
\begin{minipage}[t]{0.12\columnwidth}\centering\strut
48
\strut\end{minipage}\tabularnewline
\begin{minipage}[t]{0.19\columnwidth}\centering\strut
\textbf{Acc. t\textgreater{}1}
\strut\end{minipage} &
\begin{minipage}[t]{0.09\columnwidth}\centering\strut
\emph{86}
\strut\end{minipage} &
\begin{minipage}[t]{0.10\columnwidth}\centering\strut
68
\strut\end{minipage} &
\begin{minipage}[t]{0.12\columnwidth}\centering\strut
85
\strut\end{minipage} &
\begin{minipage}[t]{0.08\columnwidth}\centering\strut
62
\strut\end{minipage} &
\begin{minipage}[t]{0.12\columnwidth}\centering\strut
62
\strut\end{minipage}\tabularnewline
\begin{minipage}[t]{0.19\columnwidth}\centering\strut
\textbf{LL t=1}
\strut\end{minipage} &
\begin{minipage}[t]{0.09\columnwidth}\centering\strut
\emph{-656}
\strut\end{minipage} &
\begin{minipage}[t]{0.10\columnwidth}\centering\strut
-668
\strut\end{minipage} &
\begin{minipage}[t]{0.12\columnwidth}\centering\strut
-846
\strut\end{minipage} &
\begin{minipage}[t]{0.08\columnwidth}\centering\strut
-748
\strut\end{minipage} &
\begin{minipage}[t]{0.12\columnwidth}\centering\strut
-761
\strut\end{minipage}\tabularnewline
\begin{minipage}[t]{0.19\columnwidth}\centering\strut
\textbf{LL t\textgreater{}1}
\strut\end{minipage} &
\begin{minipage}[t]{0.09\columnwidth}\centering\strut
\emph{-1624}
\strut\end{minipage} &
\begin{minipage}[t]{0.10\columnwidth}\centering\strut
-2945
\strut\end{minipage} &
\begin{minipage}[t]{0.12\columnwidth}\centering\strut
-1726
\strut\end{minipage} &
\begin{minipage}[t]{0.08\columnwidth}\centering\strut
-3108
\strut\end{minipage} &
\begin{minipage}[t]{0.12\columnwidth}\centering\strut
-3146
\strut\end{minipage}\tabularnewline
\begin{minipage}[t]{0.19\columnwidth}\centering\strut
\textbf{Cor-Time}
\strut\end{minipage} &
\begin{minipage}[t]{0.09\columnwidth}\centering\strut
\emph{0.755}
\strut\end{minipage} &
\begin{minipage}[t]{0.10\columnwidth}\centering\strut
0.709
\strut\end{minipage} &
\begin{minipage}[t]{0.12\columnwidth}\centering\strut
0.713
\strut\end{minipage} &
\begin{minipage}[t]{0.08\columnwidth}\centering\strut
0.241
\strut\end{minipage} &
\begin{minipage}[t]{0.12\columnwidth}\centering\strut
-0.697
\strut\end{minipage}\tabularnewline
\begin{minipage}[t]{0.19\columnwidth}\centering\strut
\textbf{Cor-Avg.}
\strut\end{minipage} &
\begin{minipage}[t]{0.09\columnwidth}\centering\strut
0.774
\strut\end{minipage} &
\begin{minipage}[t]{0.10\columnwidth}\centering\strut
\emph{0.819}
\strut\end{minipage} &
\begin{minipage}[t]{0.12\columnwidth}\centering\strut
0.721
\strut\end{minipage} &
\begin{minipage}[t]{0.08\columnwidth}\centering\strut
0.106
\strut\end{minipage} &
\begin{minipage}[t]{0.12\columnwidth}\centering\strut
-0.724
\strut\end{minipage}\tabularnewline
\begin{minipage}[t]{0.19\columnwidth}\centering\strut
\textbf{RMSE-Time}
\strut\end{minipage} &
\begin{minipage}[t]{0.09\columnwidth}\centering\strut
\emph{0.149}
\strut\end{minipage} &
\begin{minipage}[t]{0.10\columnwidth}\centering\strut
0.154
\strut\end{minipage} &
\begin{minipage}[t]{0.12\columnwidth}\centering\strut
0.163
\strut\end{minipage} &
\begin{minipage}[t]{0.08\columnwidth}\centering\strut
0.213
\strut\end{minipage} &
\begin{minipage}[t]{0.12\columnwidth}\centering\strut
0.224
\strut\end{minipage}\tabularnewline
\begin{minipage}[t]{0.19\columnwidth}\centering\strut
\textbf{RMSE-Avg.}
\strut\end{minipage} &
\begin{minipage}[t]{0.09\columnwidth}\centering\strut
0.126
\strut\end{minipage} &
\begin{minipage}[t]{0.10\columnwidth}\centering\strut
\emph{0.113}
\strut\end{minipage} &
\begin{minipage}[t]{0.12\columnwidth}\centering\strut
0.136
\strut\end{minipage} &
\begin{minipage}[t]{0.08\columnwidth}\centering\strut
0.194
\strut\end{minipage} &
\begin{minipage}[t]{0.12\columnwidth}\centering\strut
0.203
\strut\end{minipage}\tabularnewline
\bottomrule
\end{longtable}

\textbf{Table 3.} Comparison of model performance. Best performance for
each test is \emph{italicized}. \textbf{First four rows} are performance
on 32,614 predictions of period one actions and 135,772 predictions of
period greater than one actions. Each evaluation is an average for how
that model performed with out-of-sample predictions for each game
structure. We conduct paired sample t-tests (not assuming equal
variances) to determine if the thirty accuracy and likelihood values for
the full model are statistically greater than the values of the next
best model. Accuracies for \emph{t\textgreater{}1} of the full model
(\emph{p = 0.03}) and the likelihoods for \emph{t\textgreater{}1} of the
full model (\emph{p \textless{} 0.001}) are significantly higher than
the next best model (dynamic). Accuracies for \emph{t=1} of the full
model are greater than the next best model, the static model (\emph{p =
0.07}), while the likelihoods for \emph{t=1} of the full model are not
significantly greater than the likelihoods of the static model (\emph{p
= 0.31}). \textbf{Last four rows} are performance on average cooperation
level in each structure (\emph{n=30}) and time series of average
cooperation in each structure (\emph{n=212}). Infinitely repeated
interactions with delta set to 0.5 are on average only two periods long
and there is not sufficient empirical data to extend out to eight
periods so we extend to seven. Two structures are finitely repeated for
two periods and two others are finitely repeated for four periods. We
conducted paired sample t-tests between the full model and competitors,
with a null hypothesis that the true difference in means of the 212
squared errors between predicted and real cooperation levels at all
times in all game structures is equal to zero, i.e.~that the full model
and a competitor are statistically indistinguishable in terms of squared
errors on time series predictions. We did the same for the thirty
predictions of overall cooperation levels. We reject the null of no
difference for all comparisons except with the static model for both
tests and the dynamic model for the time series (see \emph{S1
Appendix}).

Estimating the parameters of the model on a subset of the data and then
evaluating the performance of the model on held-out data allows us to
measure generalizability. However, randomly dividing the data increases
bias of the evaluation of the predictive performance because the
estimated value of the predictive power is conditional on which data
were included in the training or test samples. To reduce this bias, it
is common to run multiple rounds of this process and then average the
resulting values of predictive performance {[}41{]}. If we do this
\emph{n} times, this is called leave-out-one-cross-validation (LOOCV),
which has lower bias; however, LOOCV can have higher variance in the
estimates compared to \emph{k}-fold validation, where \(k < n\)
{[}42{]}. Fig. 5 displays the effect of the number of folds in
cross-validation on model performance, demonstrating that our main
results are robust to the value of \emph{k}.

\includegraphics{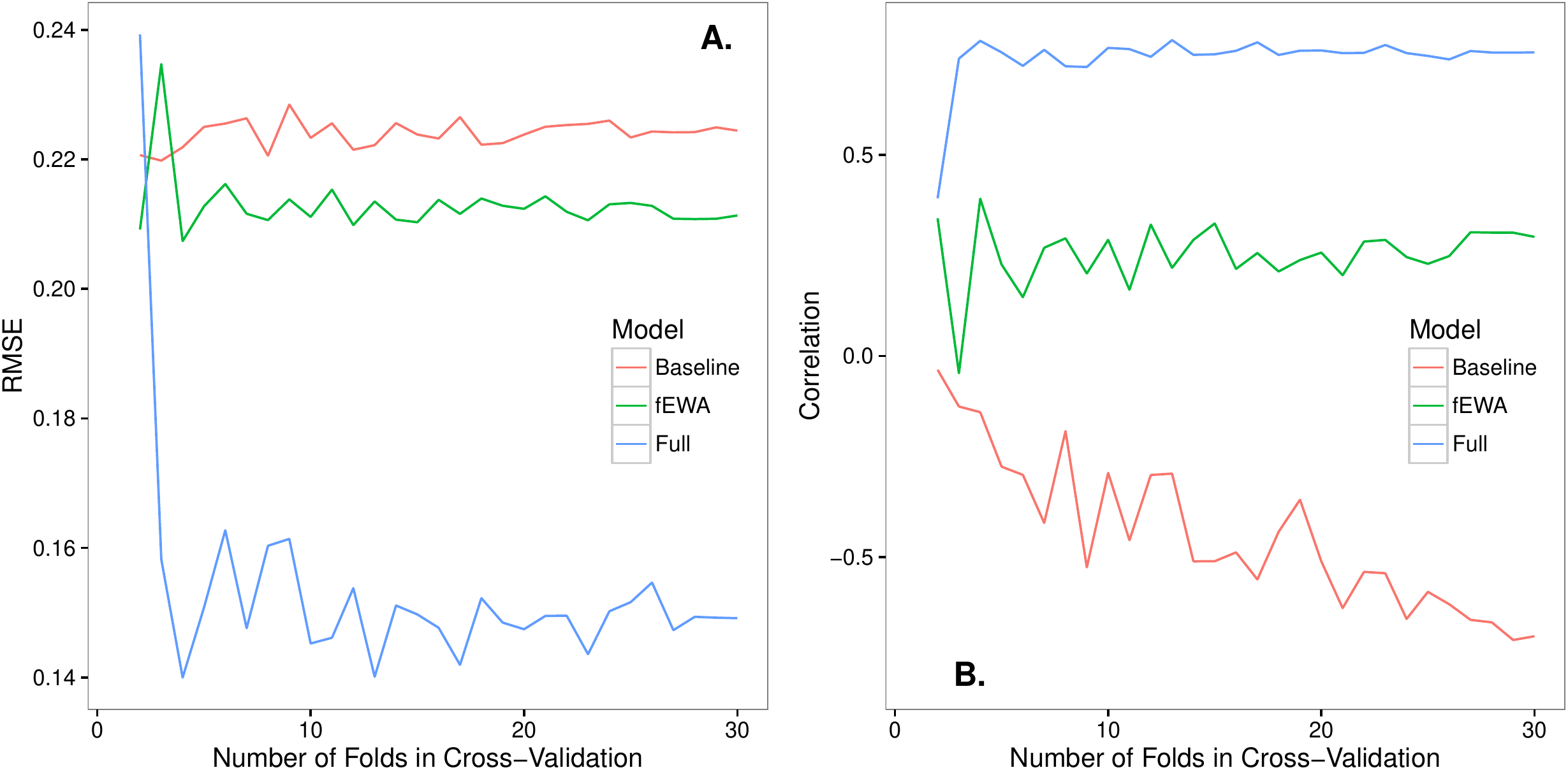}

\textbf{Fig. 5.} RMSE (\textbf{A.}) and correlation (\textbf{B.}) for
time series forecasts of play in 30 game structures, varying folds in
cross-validation from 30 to 2. The full model consistently has lower
prediction error and higher correlation than the baseline model and the
fEWA model until there are only two folds. It is, in general, difficult
to make accurate predictions when the ratio of observational units to
folds is small. In the case of predicting aggregate and dynamic play,
the game structure itself is the observational unit, and we only have
thirty, so it's not surprising that performance can degrade at two folds
depending on the particular random realization of fold assignments.

Every panel in Fig. 6 is the full model's \emph{out-of-sample forecast}
for the average probability of cooperation at each time, conditional
only on the game structure of that experiment. Our model's time series
of average cooperation is statistically significantly positively
correlated (0.76, \emph{p\textless{}0.001}) with the observed time
series. To better understand Fig. 6, observe, for example, Structure 14:
using no data from that game structure, our model predicted the initial
(high) level of cooperation almost exactly and then was perfectly
correlated with the empirically observed mean cooperation level
throughout the next seven periods of play. The \emph{S1 Appendix}
displays the equivalent of Fig. 6 for all other models, which are
noticeably worse at predicting the time series.

\includegraphics{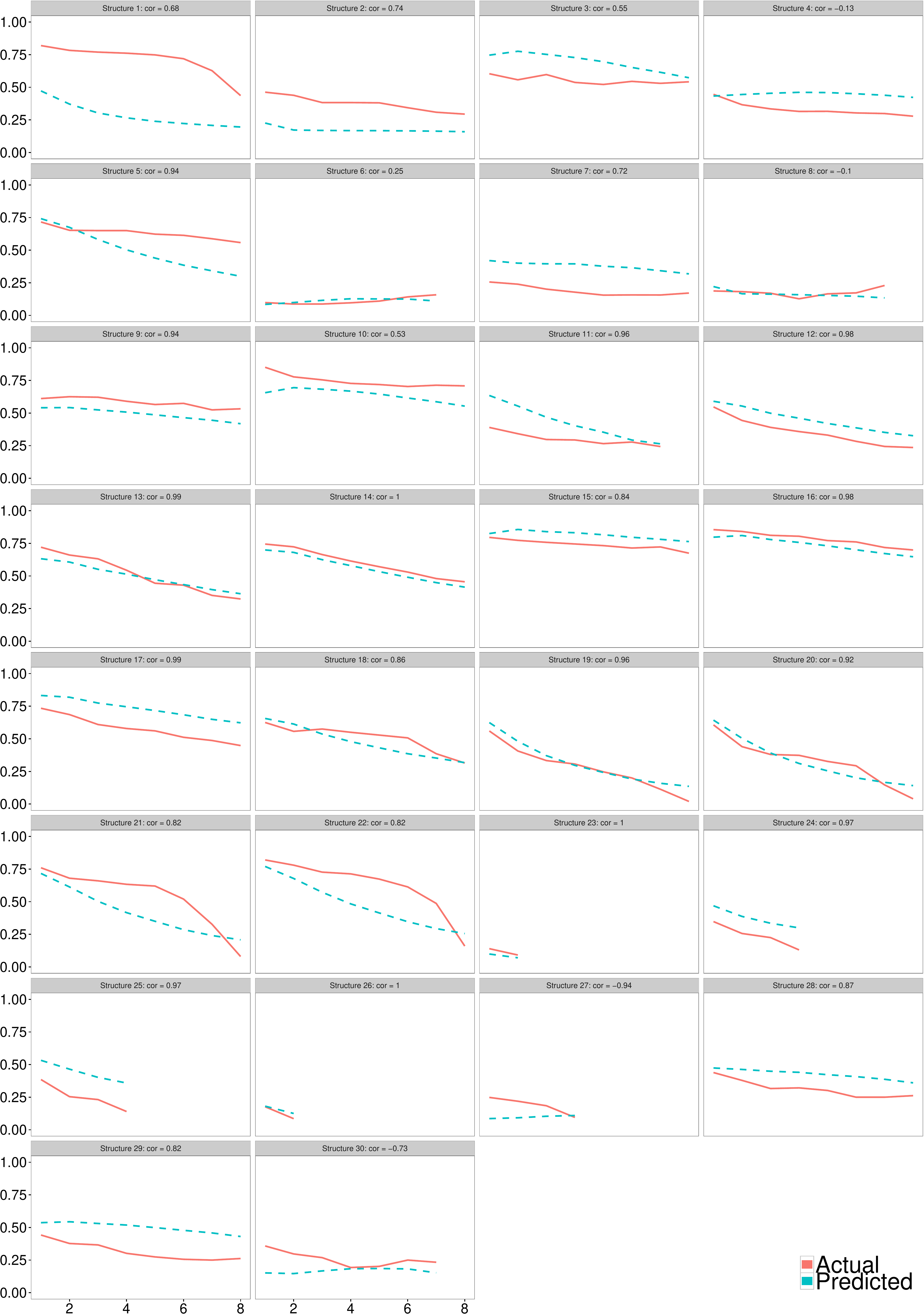}

\textbf{Fig. 6.} Out-of-sample forecasts of cooperation level over time,
for all game structures, conditional only on the game structure (\emph{n
= }212).

The dynamic-only model performed nearly as well as the full model on
individual-level \(period>1\) actions, but worse on both tests of
aggregate pattern predictions. By investigating the coefficients of the
estimated individual-level dynamic model, we discover that the actions
taken by a player and her opponent in the previous period are highly
predictive of the next action (Fig. 7). The variable with the most
predictive power is the player's own previous action: if a player
cooperated (defected) in the previous period, she is very likely to
cooperate (defect) in the next. There is strong inertia to Prisoner
Dilemma behavior, and, therefore, accurate prediction of first period
play is crucial for good performance at the aggregate level. fEWA can
incorporate the payoff game structure variables but not the other
variables, which prevents high first period accuracy. The full model is
able to predict first period play well with a model trained only on
first periods in the training data, and then use a dynamic model trained
on periods \(> 1\) in the training data, allowing for subtly different
relationships between game structures and the evolution of cooperation.

\includegraphics{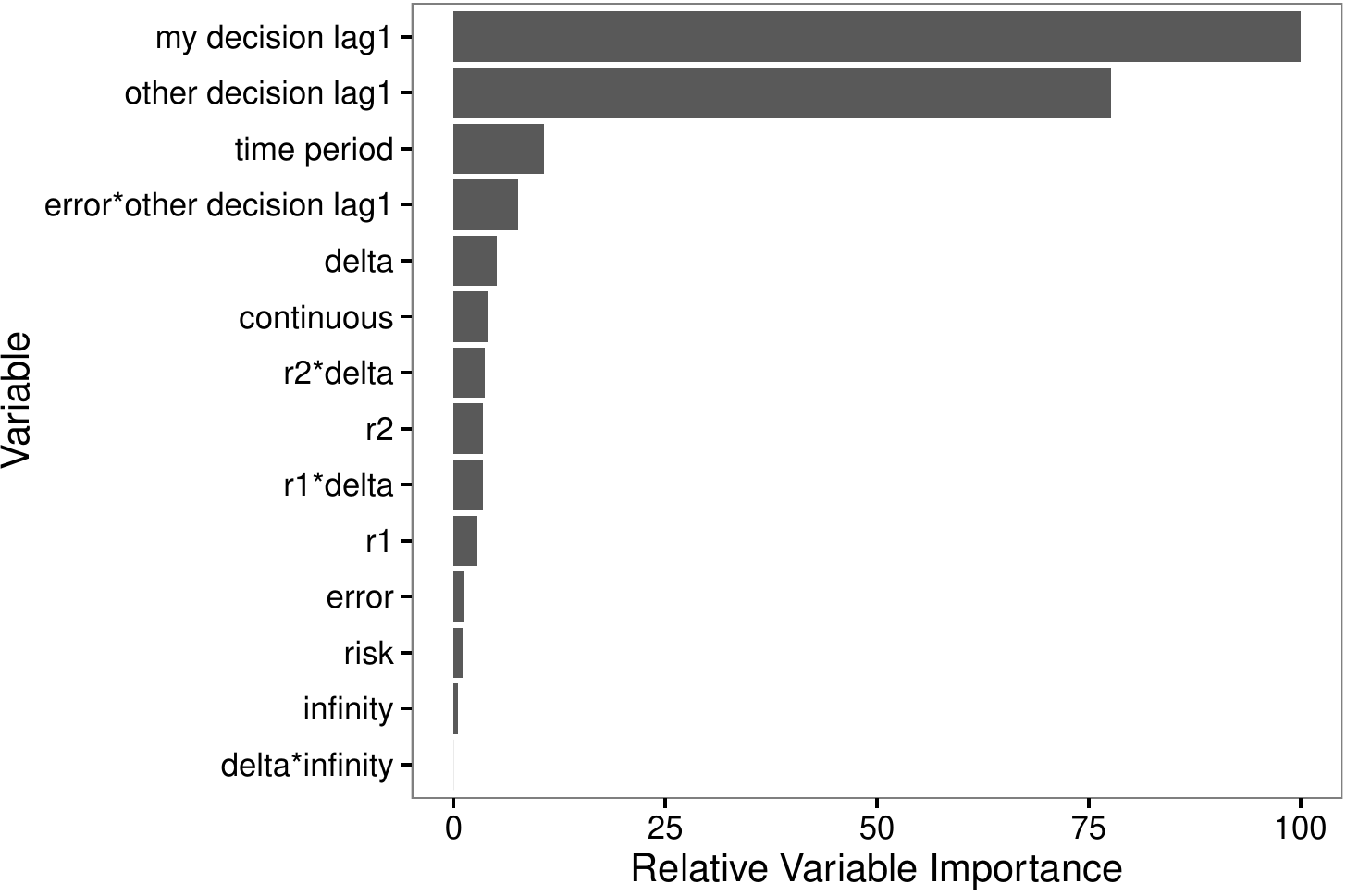}

\textbf{Fig. 7.} Variable importance scores for individual-level dynamic
component of full model, i.e.~for predictions of an agents' probability
of cooperation in periods \emph{\textgreater{} 1}. Variables separated
by `*' represent an interaction between those two variables. These
relative importance scores are derived from the absolute values of the
t-statistics for each model parameter, which correspond to the effects
of the predictor variables (accounting for variability in the estimates)
on the probability of cooperation, \emph{ceteris paribus} {[}19{]}.

The empirical experiments varied structural game parameters to measure
hypothesized differences in cooperation levels between structures. As a
final validation, we compared the (out-of-sample) predicted average
cooperation levels between our synthetic model of behavior to the actual
observed behavior in experiments {[}7,8,10--13{]}. Overall, our model
came to the same qualitative conclusions as the experiments: \(\delta\),
infinity and particular payoff configurations increased cooperation,
while risk reduced cooperation. We detail each paper's finding and
illustrate our model's corresponding finding graphically in Fig. 8.

\includegraphics{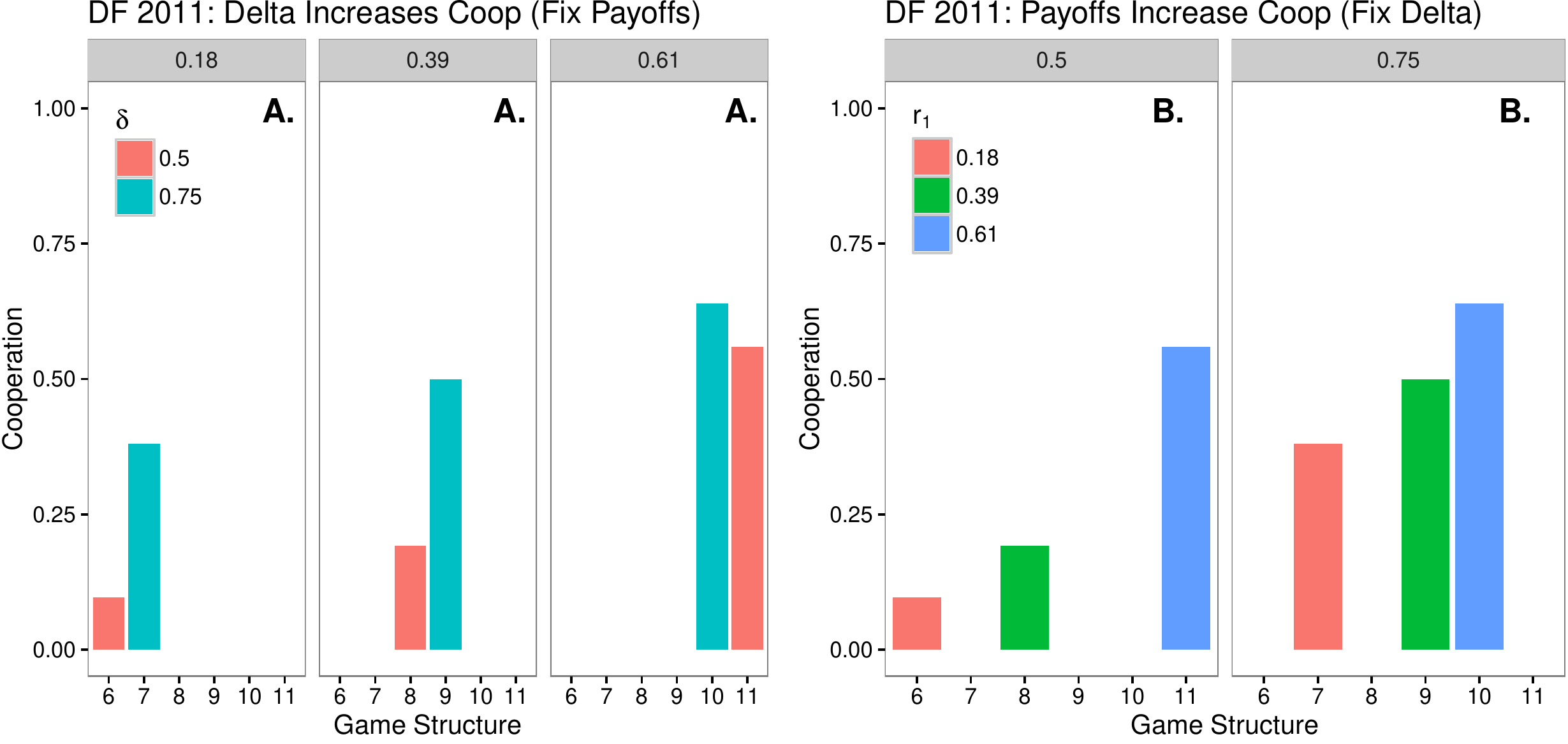}
\includegraphics{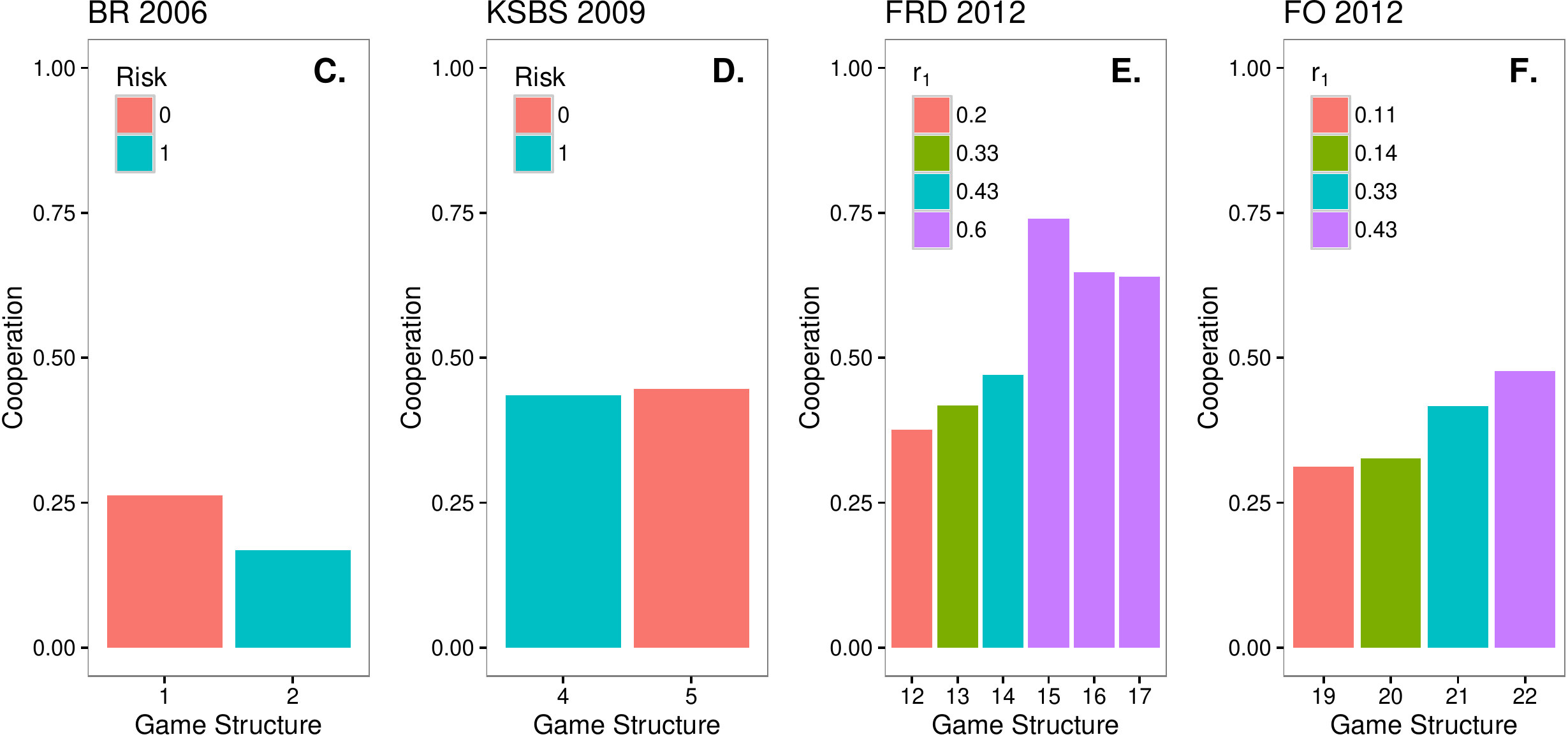}
\includegraphics{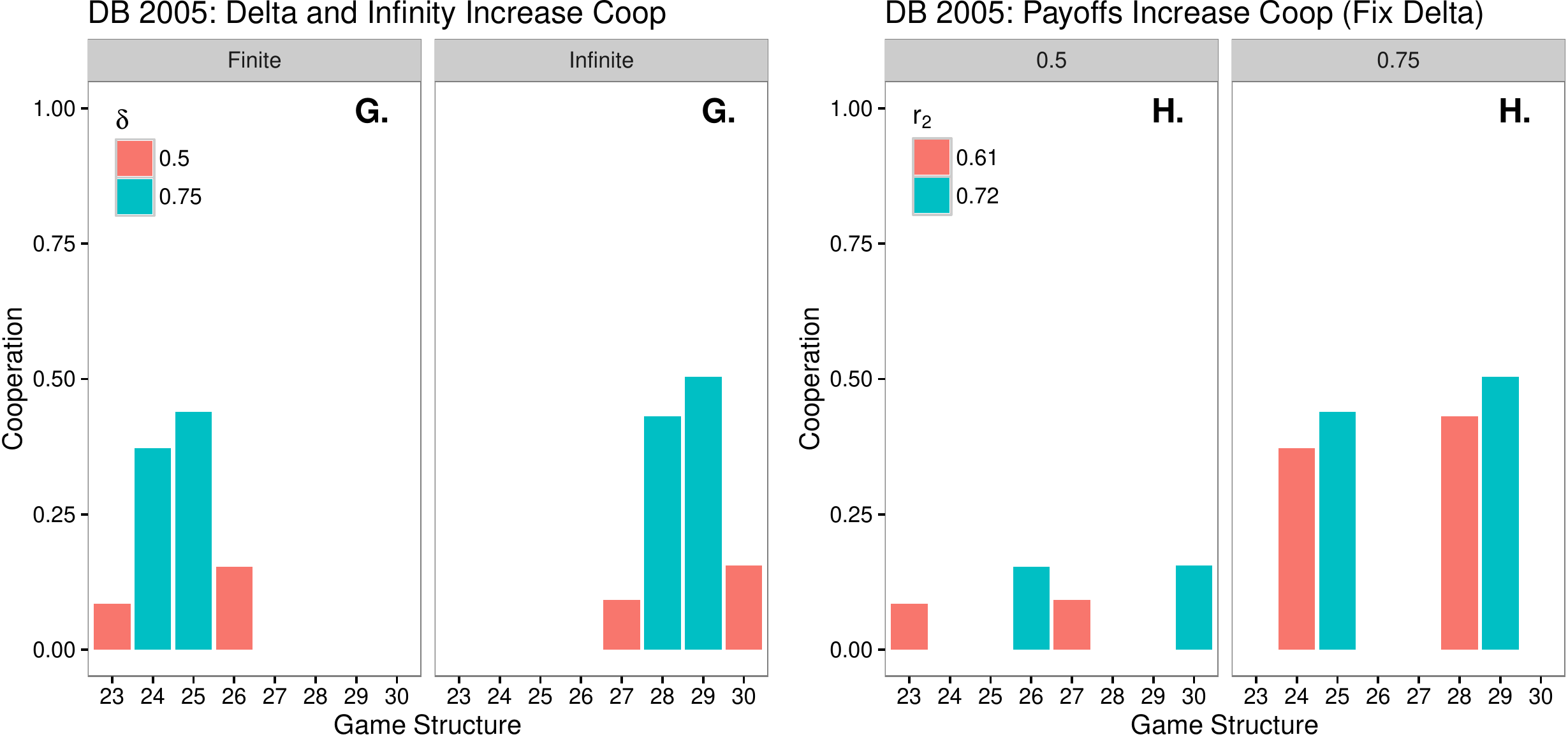}

\textbf{Fig. 8.} Predicted proportions of cooperation (\emph{n = 28}).
We could not include more than one game structure from two papers (game
structures 3 and 18) that comprised our integrated data set {[}6,9{]},
because they were comparing either to one-shot games or games with
artificial opponents. Therefore, in our model's replication of the
qualitative empirical experimental findings, we could not conduct any
replication related to these two papers' findings. Dal Bo and Frechette
found that: delta increases cooperation, keeping payoffs fixed
(\textbf{A.}); and that certain payoffs increase cooperation, while
fixing delta (\textbf{B.}). Bereby-Meyer and Roth found that risk
reduces cooperation, where payoffs were framed as gains (\textbf{C.}).
Kunreuther et al. found that risk reduces cooperation, with payoffs
framed as losses rather than gains (\textbf{D.}). This is the only
finding where we predicted marginally different cooperation levels when
the empirical data indicates a larger gap. Fudenberg Rand and Dreber
found that certain payoffs increase cooperation (\textbf{E.}). Friedman
and Oprea found that certain payoffs increase cooperation (\textbf{F.}).
Dal Bo found that delta increases cooperation, fixing payoffs and
infinity (\textbf{G.}); having an `infinitely' repeated game increases
cooperation, fixing payoffs and delta (\textbf{G.}); and certain payoffs
increase cooperation, fixing infinity and delta (\textbf{H.}). Dal Bo
also found that the cooperation levels decrease more over time within
finite games (Fig. 6 Structures 23 - 26), compared to infinite games
(Fig. 6 Structures 27 - 30).

\section{Analysis}\label{analysis}

After re-learning the computational model with all available data to
best explore the full parameter space, we deployed it to quantify the
sensitivity of cooperation to each of the structural game design
parameters. To systematically explore the model, we generated thousands
of collections of input values (specifications of Prisoner's Dilemma
experiments) from the multi-dimensional distribution covering the
feasible ranges of all input values using Latin Hypercube sampling
{[}43,44{]}. The variables are drawn from the following distributions,
with the constraint that \(r_1 < r_2\) because \(r_1\) is always less
than \(r_2\) in the data: \(error \sim Unif(0, 0.5)\);
\(\delta \sim Unif(0.45, 0.95)\); \(infinity \sim Bern(0.5)\);
\(risk \sim Bern(0.5)\); \(r_1 \sim Unif(0,1)\); \(r_2 \sim Unif(0,1)\).
Then we simulated cooperation dynamics for each experimental input set.
This global sampling and simulation allows subsequent analysis to
generate reliable information about the relationships between model
inputs (structural game design parameters) and output (cooperation
behavior) {[}45,46{]}.

Based on the results of a partial rank correlation coefficient analysis
{[}45,47{]}, the six main game structure variables can be divided into
three groups that contain two variables each within the 95\% confidence
interval of each other (Fig. 9A); we obtain qualitatively equivalent
results with a standardized rank regression coefficient analysis (see
\emph{S1 Appendix}). \(\delta\) and \(r_2\) have \emph{very large
positive effects} on average cooperation levels. As noted above and
explained further in the \emph{S1 Appendix}, our \(\delta\) measure is
applicable to both infinite and finite games as a measure of the
expected length of the game from a first period perspective, and the
dynamic model has an interaction term between \(\delta\) and
\emph{infinity} that allows the \(\delta\) effect in periods greater
than one to be different for infinite games. Surprisingly, this
interaction term is the least important predictor variable in the
dynamic model (Fig. 7), suggesting that the effect of the expected
length of the game from a first period perspective is independent of
whether the game is indefinitely repeated.

\emph{Infinity} and \(r_1\) have \emph{moderately large positive
effects} on cooperation. \(r_1\) is generally used as an index of the
cooperativeness of the payoff table so it is surprising that \(r_2\) has
a significantly larger impact on cooperation. Our analysis suggests that
we can increase the probability of cooperation more by increasing the
difference between the potential outcomes of a player and her opponent
both cooperating (C,C) and only her cooperating (C,D). Increasing the
difference between mutual cooperation (C,C) and mutual defection (D,D)
will also increase cooperation, but less. The third group includes
\emph{error} and \emph{risk}, which have \emph{negative effects} on
cooperation.

We empirically discovered that if a player cooperated (defected) in the
previous period, she is very likely to cooperate (defect) in the next
(Fig. 7). To explore the implications of this finding, we modified our
simulation model so that we could exogenously set the probability of an
agent cooperating in the first period, and found that it strongly
affects cooperation levels in subsequent periods with the game structure
set to the empirical mean values (Fig. 9B Simulated Experiments 1 and
4). However, a game structure that the analysis indicates is very
favorable to cooperation can moderate the negative effect of initial
defection (Fig. 9B Experiment 2), and, conversely, a game structure that
the analysis suggests should inhibit cooperation can moderate the
positive effect of initial cooperation (Fig. 9B Experiment 3). The
history of a particular interaction \emph{and} the institutional
structure both play important roles in determining cooperation levels.

\includegraphics{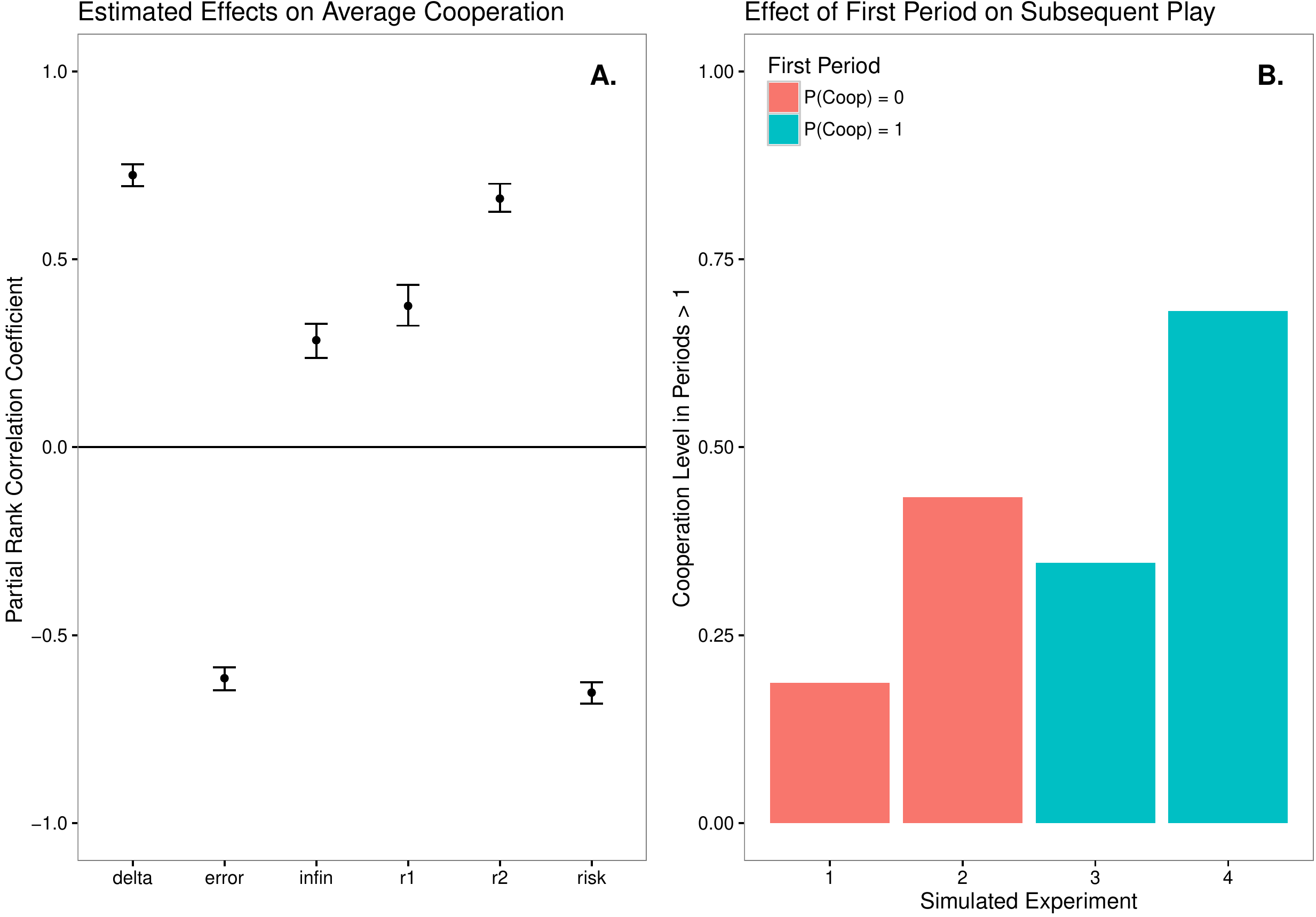}

\textbf{Fig. 9.} Model simulation analysis. \textbf{A.} is a partial
rank correlation coefficient analysis {[}47{]} of the effects of the
game parameters on average cooperation; lines are bootstrapped 95\%
confidence intervals (\emph{n = 1,000}). Continuous is set to its
empirical mode, 0, because we had no within experiment variation on
this. \textbf{B.} shows that first period play strongly affects
cooperation levels in periods greater than one. Setting the game
structure variables to the mean of the empirically observed values: if
we exogenously set the probability of cooperation during the first
period to 0 the simulated proportion of cooperation in subsequent
periods is only 0.18 (`simulated experiment 1'), and if we set the
probability of cooperation during the first period to 1 the simulated
proportion of cooperation in subsequent periods is 0.68 (`simulated
experiment 4'). When the probability of first period cooperation is set
to 0, and we use a game structure that \textbf{A.} suggests should
maximize cooperation, the proportion of cooperation is 0.43 (`simulated
experiment 2'); and when the probability of first period cooperation is
set to 1, with the game structure that should minimize cooperation, the
cooperation level is 0.35 (`simulated experiment 3').

We further investigated ``inertia'' -- the probability a player will
cooperate given that she cooperated last period -- and its relationship
to game structure. We compute an average ``predicted inertia'' for each
of the thirty game structures by predicting the probability of
cooperation after cooperating last period in a given game structure with
our model, marginalizing out the effect of the time period and the
opponent's previous decision. To compute an average ``actual inertia''
value for each of the thirty game structures we divide the sum of the
number of times all players cooperated after cooperating in the previous
period by the total number of times all players cooperated in the
previous period. The thirty predicted and actual inertia values have a
0.74 correlation, further evidence that the model captures the relevant
patterns in the data. Game structures with longer expected length of
interactions from a first-period perspective (higher \(\delta\)),
indefinite repetition, and higher \(r_2\) payoff values have higher
actual inertia values. \(\delta\) is the strongest predictor of higher
inertia and they are correlated at the 0.71 level.

\section{Conclusion}\label{conclusion}

The Prisoner's Dilemma game is widely used to understand the tension
between social and individual interests. We develop a computational
model that can accurately predict human behavior in Prisoner's Dilemma
experimental games for a broad range of game structures, using only a
few such structures for calibrating the model. We demonstrate that our
approach can successfully predict behavior at multiple scales, yielding
the most rigorously and broadly validated computational framework to
date for designing institutions that promote cooperation in social
dilemma scenarios. In particular, we use our model to identify variables
that have the greatest impact on cooperation.

Our sensitivity analysis demonstrated the importance of higher expected
values of interaction length and larger differences between potential
\emph{C,C} and \emph{C,D} outcomes (Fig. 9). It is more important to
increase the benefits of mutual cooperation over losing out by being the
sole cooperator than it is to increase the potential benefits of mutual
cooperation relative to mutual defection. These insights are relevant to
improving the underlying structure of new policy programs and designing
new human subjects experiments. This work represents a new approach to
understanding and \emph{predicting} human interactions that will be
increasingly relevant as more (experimental and observational)
behavioral data is collected. With sufficient behavioral data from a
variety of policy structures, our computational method can be applied to
understand which factors should be prioritized to improve policy
outcomes. The specifics need to be tailored to the circumstance, but
models like ours can serve as a starting point for understanding which
structural factors of a policy are most influential.

\section{Acknowledgements}\label{acknowledgements}

Nay thanks Jonathan Gilligan for discussions that improved this paper.
We thank {[}6{]}, {[}7{]}, {[}8{]}, {[}9{]}, {[}10{]}, {[}11{]},
{[}12{]}, and {[}13{]} for making their data publicly available, and
Howard Kunreuther for comments on this paper.

S1 Appendix. Supplementary Information.

\section*{References}\label{references}
\addcontentsline{toc}{section}{References}

\hypertarget{refs}{}
\hypertarget{ref-duux5fevolutionaryux5f2009}{}
1. Du W-B, Cao X-B, Zhao L, Hu M-B. Evolutionary games on scale-free
networks with a preferential selection mechanism. Physica A: Statistical
Mechanics and its Applications. 2009;388: 4509--4514.
doi:\href{https://doi.org/10.1016/j.physa.2009.07.012}{10.1016/j.physa.2009.07.012}

\hypertarget{ref-wangux5fspatialux5f2012}{}
2. Wang J, Xia C, Wang Y, Ding S, Sun J. Spatial prisoner's dilemma
games with increasing size of the interaction neighborhood on regular
lattices. Chin Sci Bull. 2012;57: 724--728.
doi:\href{https://doi.org/10.1007/s11434-011-4890-4}{10.1007/s11434-011-4890-4}

\hypertarget{ref-xiaux5fheterogeneousux5f2015}{}
3. Xia C-Y, Meng X-K, Wang Z. Heterogeneous coupling between
interdependent lattices promotes the cooperation in the prisoner's
dilemma game. PLoS One. 2015;10.
doi:\href{https://doi.org/10.1371/journal.pone.0129542}{10.1371/journal.pone.0129542}

\hypertarget{ref-hardinux5ftragedyux5f1968}{}
4. Hardin G. The tragedy of the commons. Science. 1968;162: 1243--1248.
doi:\href{https://doi.org/10.1126/science.162.3859.1243}{10.1126/science.162.3859.1243}

\hypertarget{ref-janssenux5flabux5f2010}{}
5. Janssen MA, Holahan R, Lee A, Ostrom E. Lab experiments for the study
of social-ecological systems. Science. 2010;328: 613--617.
doi:\href{https://doi.org/10.1126/science.1183532}{10.1126/science.1183532}

\hypertarget{ref-andreoniux5frationalux5f1993}{}
6. Andreoni J, Miller JH. Rational cooperation in the finitely repeated
prisoner's dilemma: Experimental evidence. The Economic Journal.
1993;103: 570--585.
doi:\href{https://doi.org/10.2307/2234532}{10.2307/2234532}

\hypertarget{ref-dalux5fboux5fcooperationux5f2005}{}
7. Dal Bo P. Cooperation under the shadow of the future: Experimental
evidence from infinitely repeated games. American Economic Review.
2005;95: 1591--1604.
doi:\href{https://doi.org/10.1257/000282805775014434}{10.1257/000282805775014434}

\hypertarget{ref-bereby-meyerux5fspeedux5f2006}{}
8. Bereby-Meyer Y, Roth AE. The speed of learning in noisy games:
Partial reinforcement and the sustainability of cooperation. American
Economic Review. 2006;96: 1029--1042. Available:
\url{http://www.ingentaconnect.com/content/aea/aer/2006/00000096/00000004/art00006}

\hypertarget{ref-duffyux5fcooperativeux5f2009}{}
9. Duffy J, Ochs J. Cooperative behavior and the frequency of social
interaction. Games and Economic Behavior. 2009;66: 785--812.
doi:\href{https://doi.org/10.1016/j.geb.2008.07.003}{10.1016/j.geb.2008.07.003}

\hypertarget{ref-kunreutherux5fbayesianux5f2009}{}
10. Kunreuther H, Silvasi G, Bradlow ET, Small D. Bayesian analysis of
deterministic and stochastic prisoner's dilemma games. Judgment and
Decision Making. 2009;4: 363--384.

\hypertarget{ref-dalux5fboux5fevolutionux5f2011}{}
11. Dal Bo P, Frechette GR. The evolution of cooperation in infinitely
repeated games: Experimental evidence. American Economic Review.
2011;101: 411--429.
doi:\href{https://doi.org/10.1257/aer.101.1.411}{10.1257/aer.101.1.411}

\hypertarget{ref-friedmanux5fcontinuousux5f2012}{}
12. Friedman D, Oprea R. A continuous dilemma. American Economic Review.
2012;102: 337--363. Available:
\url{http://www.ingentaconnect.com/content/aea/aer/2012/00000102/00000001/art00011}

\hypertarget{ref-fudenbergux5fslowux5f2012}{}
13. Fudenberg D, Rand DG, Dreber A. Slow to anger and fast to forgive:
Cooperation in an uncertain world. American Economic Review. 2012;102:
720--749.
doi:\href{https://doi.org/10.1257/aer.102.2.720}{10.1257/aer.102.2.720}

\hypertarget{ref-axelrodux5ffurtherux5f1988}{}
14. Axelrod R, Dion D. The further evolution of cooperation. Science.
1988;242: 1385--1390.
doi:\href{https://doi.org/10.1126/science.242.4884.1385}{10.1126/science.242.4884.1385}

\hypertarget{ref-rapoportux5fprisonersux5f1965}{}
15. Rapoport A, Chammah AM. Prisoners dilemma: A study in conflict and
cooperation. s.l.: University of Michigan Press; 1965.

\hypertarget{ref-rothux5fequilibriumux5f1978}{}
16. Roth AE, Keith J. Equilibrium behavior and repeated play of the
prisoner's dilemma. Journal of Mathematical Psychology. 1978;17:
189--198.
doi:\href{https://doi.org/10.1016/0022-2496(78)90030-5}{10.1016/0022-2496(78)90030-5}

\hypertarget{ref-dreberux5fwhoux5f2014}{}
17. Dreber A, Fudenberg D, Rand DG. Who cooperates in repeated games:
The role of altruism, inequity aversion, and demographics. Journal of
Economic Behavior \& Organization. 2014;98: 41--55.
doi:\href{https://doi.org/10.1016/j.jebo.2013.12.007}{10.1016/j.jebo.2013.12.007}

\hypertarget{ref-rux5fcoreux5fteamux5frux5f2015}{}
18. R\textbackslash{}\_Core\textbackslash{}\_Team. R : A language and
environment for statistical computing {[}Internet{]}. Vienna, Austria: R
Foundation for Statistical Computing; 2015. Available:
\url{http://www.R-project.org/}

\hypertarget{ref-kuhnux5fcaret:ux5f2014}{}
19. Kuhn M, Weston S, Williams A, Keefer C, Engelhardt A, Cooper T, et
al. Caret: Classification and regression training {[}Internet{]}. 2014.
Available: \url{http://cran.r-project.org/web/packages/caret/index.html}

\hypertarget{ref-houx5fself-tuningux5f2007}{}
20. Ho TH, Camerer CF, Chong J-K. Self-tuning experience weighted
attraction learning in games. Journal of Economic Theory. 2007;133:
177--198.
doi:\href{https://doi.org/10.1016/j.jet.2005.12.008}{10.1016/j.jet.2005.12.008}

\hypertarget{ref-axelrodux5fcomplexityux5f1997}{}
21. Axelrod RM. The complexity of cooperation: Agent-based models of
competition and collaboration. Princeton University Press; 1997.

\hypertarget{ref-deadmanux5fmodellingux5f1999}{}
22. Deadman PJ. Modelling individual behaviour and group performance in
an intelligent agent-based simulation of the tragedy of the commons.
Journal of Environmental Management. 1999;56: 159--172.
doi:\href{https://doi.org/10.1006/jema.1999.0272}{10.1006/jema.1999.0272}

\hypertarget{ref-jagerux5fusingux5f2002}{}
23. Jager W, Janssen MA. Using artificial agents to understand
laboratory experiments of common-pool resources with real agents.
Complexity and ecosystem management: The theory and practice of
multi-agent systems. 2002; 75--102. Available:
\url{http://www.marcojanssen.info/2002_Using_artificial_agents_to_understand_laboratory_experiments_of_common_pool_resources_with_real\%20agents.pdf}

\hypertarget{ref-janssenux5flearningux5f2006}{}
24. Janssen MA, Ahn T-K. Learning, signaling, and social preferences in
public-good games. Ecology and society. 2006;11: 21. Available:
\url{http://www.cs.sfu.ca/~lshia/personal/econ/papers/janssenAhn1.pdf}

\hypertarget{ref-wendelux5fagent-basedux5f2010}{}
25. Wendel S, Oppenheimer J. An agent-based analysis of
context-dependent preferences. Journal of Economic Psychology. 2010;31:
269--284.
doi:\href{https://doi.org/10.1016/j.joep.2009.08.005}{10.1016/j.joep.2009.08.005}

\hypertarget{ref-arifovicux5findividualux5f2012}{}
26. Arifovic J, Ledyard J. Individual evolutionary learning,
other-regarding preferences, and the voluntary contributions mechanism.
Journal of Public Economics. 2012;96: 808--823.
doi:\href{https://doi.org/10.1016/j.jpubeco.2012.05.013}{10.1016/j.jpubeco.2012.05.013}

\hypertarget{ref-wunderux5fempiricalux5f2013}{}
27. Wunder M, Suri S, Watts DJ. Empirical agent based models of
cooperation in public goods games. Proceedings of the fourteenth ACM
conference on electronic commerce. ACM; 2013. pp. 891--908. Available:
\url{http://dl.acm.org/citation.cfm?id=2482586}

\hypertarget{ref-andreoniux5fauctionsux5f1995}{}
28. Andreoni J, Miller JH. Auctions with artificial adaptive agents.
Games and Economic Behavior. 1995;10: 39--64.
doi:\href{https://doi.org/10.1006/game.1995.1024}{10.1006/game.1995.1024}

\hypertarget{ref-bowerux5fmodel-basedux5f2000}{}
29. Bower J, Bunn DW. Model-based comparisons of pool and bilateral
markets for electricity. The Energy Journal. 2000;Volume21: 1--29.
Available:
\url{https://ideas.repec.org/a/aen/journl/2000v21-03-a01.html}

\hypertarget{ref-marksux5fchapterux5f2006}{}
30. Marks R. Chapter 27 market design using agent-based models. In: L.
Tesfatsion and K.L. Judd, editor. Handbook of computational economics.
Elsevier; 2006. pp. 1339--1380. Available:
\url{http://www.sciencedirect.com/science/article/pii/S1574002105020277}

\hypertarget{ref-capraroux5fmodelux5f2013}{}
31. Capraro V. A model of human cooperation in social dilemmas. PLoS
ONE. 2013;8: e72427.
doi:\href{https://doi.org/10.1371/journal.pone.0072427}{10.1371/journal.pone.0072427}

\hypertarget{ref-capraroux5fcooperativeux5f2013}{}
32. Capraro V, Venanzi M, Polukarov M, Jennings NR. Cooperative
equilibria in iterated social dilemmas. In: Vöcking B, editor.
Algorithmic game theory. Springer Berlin Heidelberg; 2013. pp. 146--158.
Available:
\url{http://link.springer.com/chapter/10.1007/978-3-642-41392-6_13}

\hypertarget{ref-rothux5flearningux5f1995}{}
33. Roth AE, Erev I. Learning in extensive-form games: Experimental data
and simple dynamic models in the intermediate term. Games and economic
behavior. 1995;8: 164--212. Available:
\url{http://www.sciencedirect.com/science/article/pii/S089982560580020X}

\hypertarget{ref-cheungux5findividualux5f1997}{}
34. Cheung Y-W, Friedman D. Individual learning in normal form games:
Some laboratory results. Games and Economic Behavior. 1997;19: 46--76.
doi:\href{https://doi.org/10.1006/game.1997.0544}{10.1006/game.1997.0544}

\hypertarget{ref-erevux5fpredictingux5f1998}{}
35. Erev I, Roth AE. Predicting how people play games: Reinforcement
learning in experimental games with unique, mixed strategy equilibria.
American Economic Review. 1998;88: 848--81. Available:
\url{https://ideas.repec.org/a/aea/aecrev/v88y1998i4p848-81.html}

\hypertarget{ref-mckelveyux5fplayingux5f2001}{}
36. McKelvey R, Palfrey T. Playing in the dark: Information, learning,
and coordination in repeated games. California Institute of Technology;
2001.

\hypertarget{ref-camererux5fbehavioralux5f2003}{}
37. Camerer CF. Behavioral game theory: Experiments in strategic
interaction. New York, N.Y. : Princeton, N.J: Princeton University
Press; 2003.

\hypertarget{ref-stahlux5faspiration-basedux5f2002}{}
38. Stahl DO, Haruvy E. Aspiration-based and reciprocity-based rules in
learning dynamics for symmetric normal-form games. Journal of
Mathematical Psychology. 2002;46: 531--553.
doi:\href{https://doi.org/10.1006/jmps.2001.1409}{10.1006/jmps.2001.1409}

\hypertarget{ref-hanakiux5flearningux5f2005}{}
39. Hanaki N, Sethi R, Erev I, Peterhansl A. Learning strategies.
Journal of Economic Behavior \& Organization. 2005;56: 523--542.
doi:\href{https://doi.org/10.1016/j.jebo.2003.12.004}{10.1016/j.jebo.2003.12.004}

\hypertarget{ref-erevux5flearningux5f2007}{}
40. Erev I, Roth AE, Slonim RL, Barron G. Learning and equilibrium as
useful approximations: Accuracy of prediction on randomly selected
constant sum games. Economic Theory. 2007;33: 29--51.
doi:\href{https://doi.org/10.1007/s00199-007-0214-y}{10.1007/s00199-007-0214-y}

\hypertarget{ref-kimux5festimatingux5f2009}{}
41. Kim J-H. Estimating classification error rate: Repeated
cross-validation, repeated hold-out and bootstrap. Computational
Statistics \& Data Analysis. 2009;53: 3735--3745.
doi:\href{https://doi.org/10.1016/j.csda.2009.04.009}{10.1016/j.csda.2009.04.009}

\hypertarget{ref-jamesux5fintroductionux5f2013}{}
42. James G, Witten D, Hastie T, Tibshirani R. An introduction to
statistical learning: With applications in r. 1st ed. 2013. Corr. 4th
printing 2014 edition. New York: Springer; 2013.

\hypertarget{ref-beachkofskiux5fimprovedux5f2002}{}
43. Beachkofski B, Grandhi R. Improved distributed hypercube sampling.
43rd AIAA/ASME/ASCE/AHS/ASC structures, structural dynamics, and
materials conference. American Institute of Aeronautics; Astronautics;
2002. Available: \url{http://arc.aiaa.org/doi/abs/10.2514/6.2002-1274}

\hypertarget{ref-carnellux5flhsux5f2012}{}
44. Carnell R. Lhs latin hypercube samples {[}Internet{]}. 2012.
Available: \url{http://cran.r-project.org/web/packages/lhs/index.html}

\hypertarget{ref-saltelliux5fsensitivityux5f2009}{}
45. Saltelli A, Chan K, Scott EM. Sensitivity analysis. 1 edition.
Chichester: Wiley; 2009.

\hypertarget{ref-thieleux5ffacilitatingux5f2014}{}
46. Thiele JC, Kurth W, Grimm V. Facilitating parameter estimation and
sensitivity analysis of agent-based models: A cookbook using NetLogo and
r. JASSS. 2014;17: 11.

\hypertarget{ref-pujolux5fsensitivity:ux5f2014}{}
47. Pujol G, Iooss B, Lemaitre AJ with contributions from P, Gilquin L,
Gratiet LL, Touati T, et al. Sensitivity: Sensitivity analysis
{[}Internet{]}. 2014. Available:
\url{http://cran.r-project.org/web/packages/sensitivity/index.html}

\end{document}